\documentclass[aip, amsmath, amssymb, reprint]{revtex4-1}

\usepackage{graphicx}
\usepackage{dcolumn}
\usepackage{bm}
\usepackage{color}
\usepackage[utf8]{inputenc}
\usepackage[T1]{fontenc}
\usepackage{mathptmx}
\usepackage{etoolbox}
\usepackage{hyperref}

\makeatletter
\def\@email#1#2{%
 \endgroup
 \patchcmd{\titleblock@produce}
  {\frontmatter@RRAPformat}
  {\frontmatter@RRAPformat{\produce@RRAP{*#1\href{mailto:#2}{#2}}}\frontmatter@RRAPformat}
  {}{}
}%
\makeatother

\begin{document}

\preprint{AIP/123-QED}

\title[Nucleation and phase transition of spherical crystals and quasicrystals]{Nucleation and phase transition of spherical crystals and quasicrystals}

\author{Tiejun Zhou}
\affiliation{Hunan Key Laboratory for Computation and Simulation in Science and Engineering, Key Laboratory of Intelligent Computing and Information Processing of Ministry of Education, School of Mathematics and Computational Science, Xiangtan University, Xiangtan, Hunan, China, 411105.}

\author{Aibei Xie}
\affiliation{Hunan Key Laboratory for Computation and Simulation in Science and Engineering, Key Laboratory of Intelligent Computing and Information Processing of Ministry of Education, School of Mathematics and Computational Science, Xiangtan University, Xiangtan, Hunan, China, 411105.}

\author{Gang Cui}
\affiliation{Wuhan Institute for Math \& AI, Wuhan University,
Wuhan, Hubei, China, 430072.}

\author{Xianbo Huang}
\affiliation{National-certified Enterprise Technology Center, Kingfa Science and Technology Co., Ltd., Guangzhou, Guangdong, China, 510663.}

\author{Li Peng}
\affiliation{National-certified Enterprise Technology Center, Kingfa Science and Technology Co., Ltd., Guangzhou, Guangdong, China, 510663.}

\author{Kai Jiang}
\email{kaijiang@whu.edu.cn}
\affiliation{Wuhan Institute for Math \& AI, Wuhan University,
Wuhan, Hubei, China, 430072.}

\date{\today}

\begin{abstract}
We investigated the nucleation, growth, and solid-to-solid phase transition processes of spherical six-fold symmetric crystals (C6) and ten-fold symmetric quasicrystals (DQC) using the Lifshitz-Petrich model. By combining the spherical-harmonic pseudospectral method with a nullspace-preserving saddle-point search technique, we resolved the critical nucleus structures and phase transition pathways. Although both transitions from the homogeneous state to the ordered state involve the formation of localized critical nuclei, their subsequent growth processes exhibit distinct modes of accommodating geometric mismatch: C6 forms branched domain structures and persistent line defects associated with orientation and phase mismatches, whereas DQC relaxes collision-induced defects through local coordination rearrangements. The resolved DQC-to-C6 pathway proceeds through a series of
metastable intermediates that progressively connect these
distinct modes of mismatch accommodation.
Its multi-step character reflects the coupled adjustment
of local coordination and compatibility between the emerging
C6 domains. These results reveal the intrinsic link between the multi-step nature of the phase transition and the distinct structural requirements for accommodating geometric mismatch.
\end{abstract}

\maketitle

\section{Introduction}

Ordered structures on spherical surfaces arise in a broad range of natural and soft-matter systems, including virus capsids, pollen exines, emulsion droplets, spherical shells, and curved crystalline assemblies~\cite{caspar1962physical,konevtsova2012chiral,lavrentovich2016first,bausch2003grain,brojan2015wrinkling}. These systems commonly exhibit crystalline order together with curvature-induced defects, as seen in hexagonally ordered capsids, particle assemblies on droplets, and active spherical crystals~\cite{caspar1962physical,paquay2017impact,bausch2003grain,praetorius2018active}. Experiments on spherical colloidal crystals have further revealed defect segregation during freezing and an intermediate hexatic phase during melting~\cite{guerra2018freezing,singh2022observation}.
Spherical quasicrystalline order has also been investigated through geometric constructions and density-wave descriptions, motivated by its relevance to viral capsids and proteinaceous nanoshells~\cite{konevtsova2012chiral,shevchenko2014dodecagonal,konevtsova2017quasicrystalline,rochal2024proteinaceous,chalin2023landau,rochal2016soft,rochal2025anomalous, konevtsova2017quasicrystalline}.
Numerical studies of cholesteric shells have also identified finite quasicrystalline tessellations and mapped their phase behavior as functions of shell radius and elastic parameters~\cite{carenza2022cholesteric,negro2023topological}.
However, the nucleation and phase-transition mechanisms involving spherical quasicrystalline order remain much less understood.
Exploring phase transitions between spherical quasicrystalline and crystalline order is therefore important for clarifying how curvature, topology, and local symmetry jointly control the structural evolution of ordered matter on closed surfaces.

Recent studies of quasicrystals have revealed several important features of quasicrystal--crystal transitions. Theoretical calculations in planar systems show that such transitions may proceed either through a direct pathway or through a multi-step cascade involving metastable intermediate states, with the latter often favored by a lower activation barrier~\cite{yin2021transition,zhou2024nucleation}. Experimentally, real-space observations of dodecagonal quasicrystal nucleation in colloidal systems have provided direct evidence for heterogeneous and homogeneous nucleation kinetics~\cite{gao2025direct}.
Numerical and experimental studies of quasicrystal growth have also shown that additional phasonic degrees of freedom can facilitate stress relaxation and grain coalescence under suitable conditions~\cite{schmiedeberg2017dislocation,han2021formation}.
Quasicrystals can also accommodate localized geometric perturbations, for example by growing around macroscopic obstacles without forming defects~\cite{wang2025defect}. However, global spherical curvature introduces a qualitatively different constraint. On a sphere, locally preferred ordering cannot generally extend over the entire surface without distortion, and even simple spherical crystals must accommodate elastic frustration through unconventional growth modes and topological defects~\cite{meng2014elastic,qiu2026exploring}.
Numerical studies have shown that curvature-induced stress can be relieved through either defect incorporation or branched ribbonlike growth~\cite{kohler2016stress,ortellado2022crystalization}. Studies of nucleation on curved surfaces have further demonstrated that geometry affects critical nucleus sizes and transition pathways, while interface fluctuations modify the nucleation energetics~\cite{gomez2015phase,horsley2018aspects}. This raises a fundamental question: how do quasicrystalline structures with high-order local rotational symmetry nucleate, grow, and transform on a closed curved surface?

Recent saddle-point calculations for the spherical Landau--Brazovskii model have revealed radius-dependent, multi-step transitions between striped and spotted phases~\cite{qiu2026exploring}. However, the corresponding critical nuclei and transition pathways involving spherical quasicrystalline order remain poorly understood. Spherical confinement creates several difficulties that are absent in flat geometries. First, maintaining locally preferred quasicrystalline arrangements while satisfying global geometric compatibility requires structural distortions and rearrangements~\cite{konevtsova2012chiral,shevchenko2014dodecagonal,rochal2016soft}. Second, topologically required defects and curvature-induced elastic frustration can influence nucleation pathways and modify equilibrium configurations~\cite{bowick2006crystalline,qiu2026exploring}. Third, standard planar tilings and numerical solvers require adaptation, because both the construction of low-energy initial states and the computation of transition states must account for the spherical geometry. These issues are particularly severe for rare-event calculations, where the critical nucleus of a quasicrystal may exhibit a complex spatial structure and the transition may involve multiple metastable intermediates. Efficient methods based on spherical representations and saddle-search algorithms are therefore required~\cite{zavodnik2025spherical,qiu2025efficient}.

In this work, we investigate the nucleation, growth, and
solid-to-solid transformation of sixfold crystals (C6) and
decagonal quasicrystals (DQC) on a spherical surface using
the Lifshitz--Petrich (LP) model~\cite{lifshitz1997theoretical}.
We construct spherical initial configurations through a
generalized polyhedral tiling projection and determine the
relative stability of the candidate states.
By combining a spherical harmonic pseudospectral method with
the nullspace-preserving saddle search (NPSS)
method~\cite{cui2025efficient}, we resolve critical nuclei and
transition pathways from the homogeneous phase (H) to C6,
from H to DQC, and from DQC to C6.
Together with bond-orientational order analysis and spherical
Voronoi tessellation, these calculations establish how local
structural order governs the accommodation of geometric
mismatch during nucleation, growth, and transformation.
Both C6 and DQC emerge through localized critical nuclei,
but their subsequent growth follows distinct mechanisms:
C6 develops branched domains and persistent line defects
associated with orientational and phase mismatches, whereas
DQC relaxes collision-induced defects through local
coordination rearrangements.
The DQC-to-C6 pathway further demonstrates that these two
modes of mismatch accommodation can be connected through
structural transformation.
Their different compatibility requirements couple local
network reconstruction to domain-boundary reorganization,
producing a multi-step pathway through metastable
intermediates.
These results provide a unified picture of how
crystalline and quasicrystalline structures form, accommodate
spherical frustration, and transform between distinct modes
of defect organization.

\section{Theoretical Model and Computational Framework}

\subsection{Spherical Lifshitz--Petrich Model}
The LP model describes quasiperiodic ordering through the competition between two characteristic length scales. It has been widely used to study the formation and stability of soft-matter quasicrystals~\cite{lifshitz2007soft,jiang2015stability}, their phase transition mechanisms~\cite{yin2021transition,zhou2024nucleation}, and defect dynamics~\cite{barak2006dislocation}. In the LP framework, the local density variation is represented by a scalar order parameter $\psi(\mathbf{r})$. To model ordered structures confined to a curved surface, we restrict the domain to a sphere of radius $R$,
\begin{equation}
\mathbb{S}^2_R=\{(x,y,z)\in\mathbb{R}^3:x^2+y^2+z^2=R^2\}.
\end{equation}
The free energy of the spherical LP model is given by
\begin{multline}\label{eq:2-1}
E(\psi) = \frac{1}{|\mathbb{S}^2_R|} \int_{\mathbb{S}^2_R} \Bigg\{
    \frac{1}{2}\big[(\Delta_{\mathbb{S}^{2}_R}+1)(\Delta_{\mathbb{S}^{2}_R}+q^2)\psi\big]^2 \\
    - \frac{\alpha}{2}\psi^2 - \frac{\gamma}{3}\psi^3 + \frac{1}{4}\psi^4
\Bigg\} \mathrm{d}s,
\end{multline}
where $\Delta_{\mathbb{S}^{2}_R}$ denotes the Laplace--Beltrami operator on the sphere, $\mathrm{d}s$ is the surface area element, and $|\mathbb{S}^2_R|=4\pi R^2$. All free energies reported below are area-averaged quantities.

Using spherical coordinates, a point on $\mathbb{S}^2_R$ is parameterized as
\begin{equation}
\mathbf{r}(\theta,\varphi)
=
(R\sin\theta\cos\varphi,\,
 R\sin\theta\sin\varphi,\,
 R\cos\theta),
\end{equation}
where $\theta\in[0,\pi]$ is the colatitude and $\varphi\in[0,2\pi)$ is the azimuthal angle. Accordingly, the order parameter can be written as $\psi(\theta,\varphi)=\psi(\mathbf{r}(\theta,\varphi))$. The parameter $\alpha$ is a temperature-like control parameter, $\gamma$ controls the strength of the cubic interaction, and $1$ and $q$ specify the two preferred wavenumbers. On the sphere, these preferred wavenumbers are realized through the discrete spectrum of the Laplace--Beltrami operator, so the radius $R$ controls the compatibility between the intrinsic LP length scales and the spherical geometry.

We impose a zero-mean constraint so that $\psi$ represents density variations relative to a uniform background:
\begin{equation}\label{eq:mass_conservation}
\frac{1}{|\mathbb{S}^2_R|}\int_{\mathbb{S}^2_R}\psi(\mathbf{r})\,\mathrm{d}s=0.
\end{equation}

For numerical simulations, Eq.~\eqref{eq:2-1} is discretized using a spherical harmonic pseudospectral method on a Gauss--Legendre grid. The mesh contains $N_\theta=256$ Gauss–Legendre colatitude nodes and $N_\varphi=512$ uniform longitude points, and the spherical harmonic expansion is truncated at degree $N=255$. Linear differential operators are evaluated in spectral space, while nonlinear terms are computed pointwise in physical space. The forward and inverse spherical harmonic transforms are performed using the open-source package SHTns~\cite{schaeffer2013efficient}. The zero-mean constraint in Eq.~\eqref{eq:mass_conservation} is enforced by projecting the discrete field onto the zero-mean subspace. Further details on the spherical harmonic representation, quadrature rule, and spectral discretization are given in Appendix~\ref{app:discretization}.

\subsection{Initial Configuration of Spherical Quasicrystals}

The classical Caspar--Klug theory, based on the principle of quasi-equivalence, provides a fundamental geometric framework for describing spherical viral capsids~\cite{caspar1962physical}. However, this theory is less suitable for capsids with nonstandard local symmetry or more complex organizational motifs. To extend the description beyond conventional crystalline order, Twarock introduced T-K tilings for viral architectures with locally distributed pentagonal symmetry~\cite{twarock2004tiling}. Inspired by the self-assembly of binary colloidal mixtures, Shevchenko \textit{et al.} further proposed a spherical dodecagonal quasicrystalline tiling generated by controlled geometric projections~\cite{shevchenko2014dodecagonal}. These studies suggest that geometric projection offers a practical route for constructing noncrystalline ordered states on spherical surfaces.

\begin{figure}[htbp]
 \centering
 \includegraphics[width=0.48\textwidth]{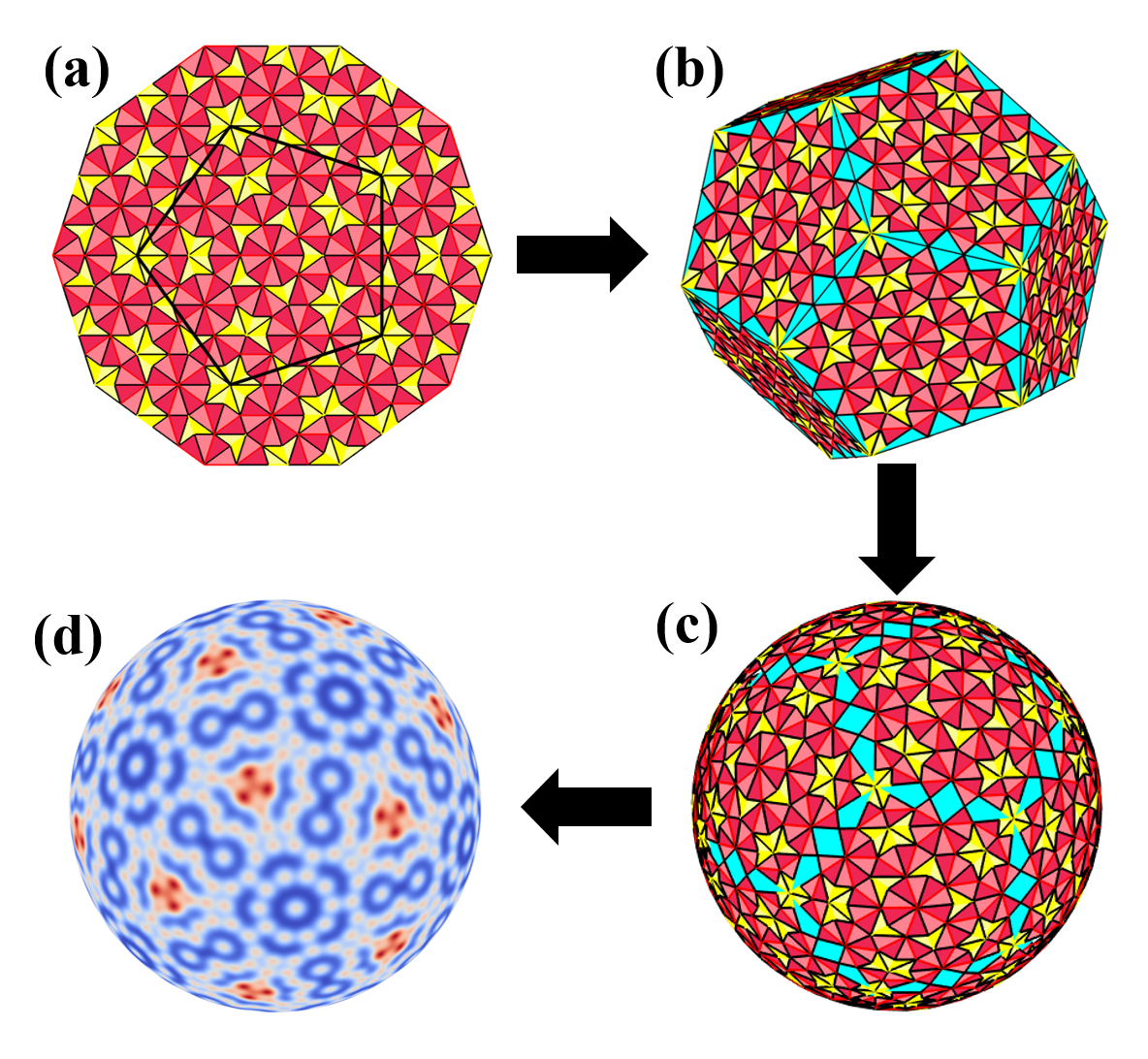}
 \caption{Construction of the initial ansatz for the spherical DQC. 
 (a) A pentagonal patch extracted from a planar P2 Penrose tiling. 
 (b) Face-wise affine mapping of the patch onto a regular dodecahedron. 
 (c) Radial projection of the polyhedral nodes onto $\mathbb{S}^2_R$. 
 (d) Smooth initial field obtained by superposing Gaussian kernels centered at the projected nodes.} 
 \label{fig:sph_qc_tiling}
\end{figure}

Motivated by these structural models, we construct an initial ansatz for the spherical DQC using a face-wise geometric projection protocol. We adopt the P2 Penrose tiling~\cite{penrose1979pentaplexity} as a convenient planar template~\cite{steurer2014decagonal}. This choice is motivated by the qualitative agreement between the local vertex arrangements of the P2 tiling and the characteristic motifs observed in decagonal quasicrystals, as illustrated in Fig.~\ref{fig:sph_qc_tiling}.

Since the P2 Penrose tiling is naturally organized around pentagonal motifs, a regular dodecahedron provides a suitable intermediate scaffold for spherical projection. Unlike an icosahedron with triangular faces, the dodecahedron contains twelve regular pentagonal faces, each of which can accommodate a fivefold-symmetric patch. The regular dodecahedron provides a scaffold with icosahedral symmetry, whereas the symmetry of the decorated node pattern depends on the choice and orientation of the patches on its faces.

The initial state is then generated through the following three steps:
\begin{enumerate}
    \item \textbf{Patch excision and affine mapping.} 
    A regular pentagonal patch is first extracted from the planar P2 Penrose tiling [Fig.~\ref{fig:sph_qc_tiling}(a)]. The patch is duplicated and mapped onto the twelve faces of a regular dodecahedron by local affine transformations, yielding a decorated polyhedral node pattern [Fig.~\ref{fig:sph_qc_tiling}(b)].

    \item \textbf{Radial projection.} 
    Let $\mathbf{x}_j$ denote the vertices of the dodecahedral node pattern. These vertices are radially projected onto the sphere of radius $R$ by
    \begin{equation}
    \mathbf{r}_j = R\frac{\mathbf{x}_j}{|\mathbf{x}_j|},
    \end{equation}
    which defines the spherical node positions [Fig.~\ref{fig:sph_qc_tiling}(c)].

    \item \textbf{Continuous field regularization.} 
    To obtain a smooth initial ansatz for the LP model, the discrete node set is regularized by a superposition of localized Gaussian kernels:
    \begin{equation}
    \psi_{\mathrm{init}}(\mathbf{r}) =
    \mathcal{P}_0
    \sum_j
    \exp\left(
    -\frac{|\mathbf{r}-\mathbf{r}_j|^2}{2\sigma^2}
    \right),
    \end{equation}
    where $|\mathbf{r}-\mathbf{r}_j|$ denotes the chordal distance in $\mathbb{R}^3$, and $\sigma$ is chosen according to the characteristic nearest-neighbor spacing. The operator $\mathcal{P}_0$ projects the field onto the zero-mean subspace,
    \begin{equation}
    \mathcal{P}_0 f
    =
    f-\frac{1}{|\mathbb{S}^2_R|}
    \int_{\mathbb{S}^2_R}f(\mathbf{r})\,\mathrm{d}s,
    \end{equation}
    so that the mass-conservation constraint is satisfied [Fig.~\ref{fig:sph_qc_tiling}(d)].
\end{enumerate}
Sampling this continuous field on the Gauss--Legendre grid gives the initial DQC configuration used in the subsequent numerical simulations.
This geometrically constructed field serves only as an initial ansatz. The final relaxed DQC state is obtained by subsequent energy relaxation under the spherical LP model.

\subsection{Phase Diagram}
In this study, we fix the second characteristic length-scale parameter at
$q=2\cos(\pi/5)$, which corresponds to the two-scale resonance associated
with decagonal ordering in the LP model. To assess the
relative thermodynamic stability of the ordered configurations, we construct
a phase diagram in the $(\alpha,\gamma)$ parameter space on a
spherical domain with radius $R=60.5$. Three representative phases are
considered: H, C6,
and DQC.

For each parameter pair $(\alpha,\gamma)$, the prescribed initial
configurations for H, C6, and DQC are relaxed to their corresponding local
minimizers under the mass-conserving gradient flow
\begin{equation}
\frac{\partial \psi}{\partial t}
=
-\nabla_{\mathcal{M}}E(\psi),
\label{eq:gradient_flow_phase_diagram}
\end{equation}
where
\begin{equation}
\mathcal{M}
=
\left\{
\psi:
\int_{\mathbb{S}^2_R}\psi(\mathbf{r})\,\mathrm{d}s=0
\right\},
\end{equation}
and $\nabla_{\mathcal{M}}E$ denotes the $L^2(\mathbb{S}^2_R)$-gradient of
$E$ projected onto the zero-mean constraint manifold. The relaxation is
terminated when the projected gradient satisfies
\begin{equation}
\|\nabla_{\mathcal{M}}E(\psi)\|_{L^2(\mathbb{S}^2_R)}<\epsilon.
\label{eq:phase_diagram_convergence}
\end{equation}
After convergence, the free energies of the relaxed H, C6, and DQC states
are compared at each parameter point. The state with the lowest value of
$E$ among these three candidates is identified as the numerically preferred
phase.

The resulting candidate-state phase diagram identifies the regions of energetic preference among the relaxed H, C6, and DQC structures and provides a reference for selecting the ordered states investigated below.

\begin{figure}[htbp]
 \centering
 \includegraphics[width=0.48\textwidth]{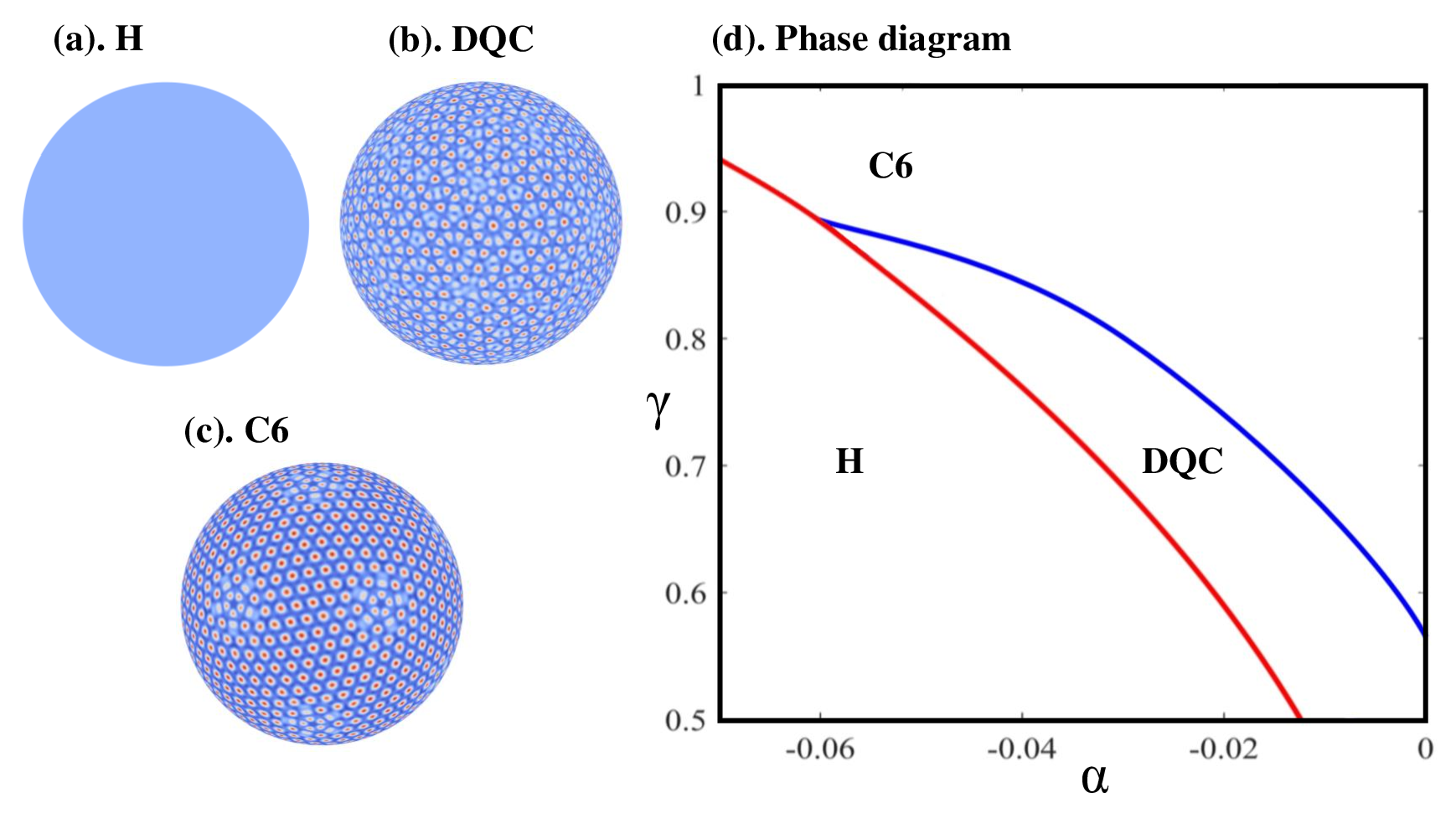}
 \caption{Spherical ordered structures and the reference phase diagram of the spherical LP model with $q=2\cos(\pi/5)$ and $R=60.5$.
 (a) H. (b) DQC.
 (c) C6. (d) Phase diagram provides a reference at R = 60.5. The transition pathways investigated below are computed at different radii, and the energetic ordering of their endpoint states is evaluated separately at the corresponding parameter values.}
 \label{fig:state}
\end{figure}

Figure~\ref{fig:state} shows the relaxed spherical structures and the
resulting candidate-state phase diagram. For the representative parameters
$\alpha=-0.03$ and $\gamma=0.8$, the DQC has a lower free energy than both the homogeneous
state and the C6 crystal, and is therefore the preferred candidate in this
regime. Compared with the continuous quasiperiodic structure of the DQC state,
the C6 lattice contains twelve prominent fivefold disclinations. These
defects are required by Euler's theorem for a triangulated spherical crystal and reflect curvature-induced geometric frustration in
periodic order on a closed surface.

\subsection{Nullspace-preserving Saddle Search Method}

Due to the continuous rotational symmetry of the sphere, spherical ordered
structures possess geometric degeneracy~\cite{cui2025efficient}. For the
spherical C6 and DQC states considered here, the degenerate nullspace is
three-dimensional in our computations. This is consistent with the continuous
$SO(3)$ rotational symmetry of the sphere: a rigid rotation changes the
orientation of the ordered pattern but leaves the free energy invariant.
Consequently, the continuous Hessian possesses three zero modes associated
with infinitesimal rotations, which appear as near-zero eigenmodes after
spatial discretization. These neutral modes do not provide effective climbing
directions. If such modes are selected as ascent directions, the saddle-search
dynamics may fail to escape from the basin of a generalized local minimum
$\psi_0$.

To overcome this difficulty, we employ the NPSS method, which was developed for ordered phase transitions involving
degeneracy~\cite{cui2025efficient}. The key idea is to constrain the ascent
direction to remain orthogonal to the degenerate nullspace, thereby excluding
neutral rotational modes from the climbing subspace. In what follows, all
gradients and Hessian-vector products are understood on the zero-mean
constraint manifold $\mathcal{M}$.

Let $\bar{\psi}$ be a reference state, initially chosen as
$\bar{\psi}=\psi_0$. We define the degenerate nullspace of the Hessian $H_{\mathcal M}(\bar\psi)$ as
\begin{equation}
\mathcal{W}(\bar{\psi})
=
\operatorname{span}\{\bar{w}_1,\bar{w}_2,\ldots,\bar{w}_l\},
\end{equation}
where
$\{\bar{w}_i\}_{i=1}^{l}$ form an orthonormal basis. The ascent direction $v$
is constrained to lie in the orthogonal complement
$\mathcal{W}(\bar{\psi})^\perp$. For a one-dimensional climbing subspace
$\mathcal{V}=\operatorname{span}\{v\}$, the state variable is evolved by
\begin{equation}
\beta^{-1}\dot{\psi}
=
-\nabla_{\mathcal{M}}E(\psi)
+
2\mathcal{P}_{\mathcal{V}}\nabla_{\mathcal{M}}E(\psi),
\label{eq:npss_state}
\end{equation}
where $\beta>0$ is a relaxation parameter and
$\mathcal{P}_{\mathcal{V}}$ denotes the orthogonal projection onto
$\mathcal{V}$. Equation~\eqref{eq:npss_state} reverses the gradient direction
inside the climbing subspace while retaining the ordinary descent direction
in its orthogonal complement.

The ascent direction is updated according to
\begin{equation}
\zeta^{-1}\dot{v}
=
-H_{\mathcal{M}}(\psi)v
+
\langle v,H_{\mathcal{M}}(\psi)v\rangle v
+
\sum_{i=1}^{l}
\langle \bar{w}_i,H_{\mathcal{M}}(\psi)v\rangle \bar{w}_i,
\label{eq:npss_direction}
\end{equation}
together with the normalization constraint $\|v\|_{L^2(\mathbb{S}^2_R)}=1$. Here $\zeta>0$ is a
relaxation parameter. The last term enforces the orthogonality of $v$ to the
fixed nullspace $\mathcal{W}(\bar{\psi})$ during each segment.

To reduce the cost of repeatedly computing degenerate eigenvectors, NPSS
uses a segmented nullspace-update strategy. Within each segment, the
reference state $\bar{\psi}$ and its nullspace $\mathcal{W}(\bar{\psi})$ are
kept fixed. When the Rayleigh quotient
$\langle v,H_{\mathcal{M}}(\psi)v\rangle$ falls below a prescribed threshold
$\epsilon_v>0$, the fixed nullspace is regarded as no longer sufficiently
accurate for the current state. The reference state is then reset to the
current state, $\bar{\psi}\leftarrow\psi$, the nullspace
$\mathcal{W}(\bar{\psi})$ is recomputed, and a new segment starts.

Once a negative-curvature direction is detected, namely
$\langle v,H_{\mathcal{M}}(\psi)v\rangle<0$, the ascent direction is updated
by the standard normalized eigenvector dynamics
\begin{equation}
\zeta^{-1}\dot{v}
=
-H_{\mathcal{M}}(\psi)v
+
\langle v,H_{\mathcal{M}}(\psi)v\rangle v.
\label{eq:standard_direction}
\end{equation}
The NPSS iteration is terminated when
\begin{equation}
\|\nabla_{\mathcal{M}}E(\psi)\|_{L^2(\mathbb{S}^2_R)}<\epsilon.
\end{equation}
We use $\varepsilon=10^{-7}$ in all saddle-search computations.
After convergence, the constrained Hessian spectrum is examined
to verify that the stationary state has exactly one negative
eigenvalue, with the rotational zero modes excluded. For a converged saddle $\psi^*$ with normalized unstable
eigenmode $v_-$, the two initial conditions
$\psi^*\pm\eta v_-$, with $\eta>0$ sufficiently small,
are relaxed using the constrained gradient flow.
The two descent branches identify the adjacent minima and define
the connecting transition-path segment.
The DQC-to-C6 pathway is assembled from successive segments
connecting the resolved metastable intermediates.

\section{Results and discussion}

\subsection{Nucleation and Growth of C6 from H}\label{H_C6}

We first examine the curvature-mediated nucleation and growth of the spherical
C6 phase from the metastable homogeneous state H. The computation is performed
on a sphere of radius $R=57$ with $\alpha=-0.002$ and $\gamma=0.8$. Under these
parameters, H is metastable with $E=0$, whereas the relaxed C6 state has a
lower area-averaged free energy, $E=-3.86\times 10^{-3}$, providing the
thermodynamic driving force for the H-to-C6 transition.

\begin{figure}[htbp]
	\centering
	\includegraphics[width=0.48\textwidth]{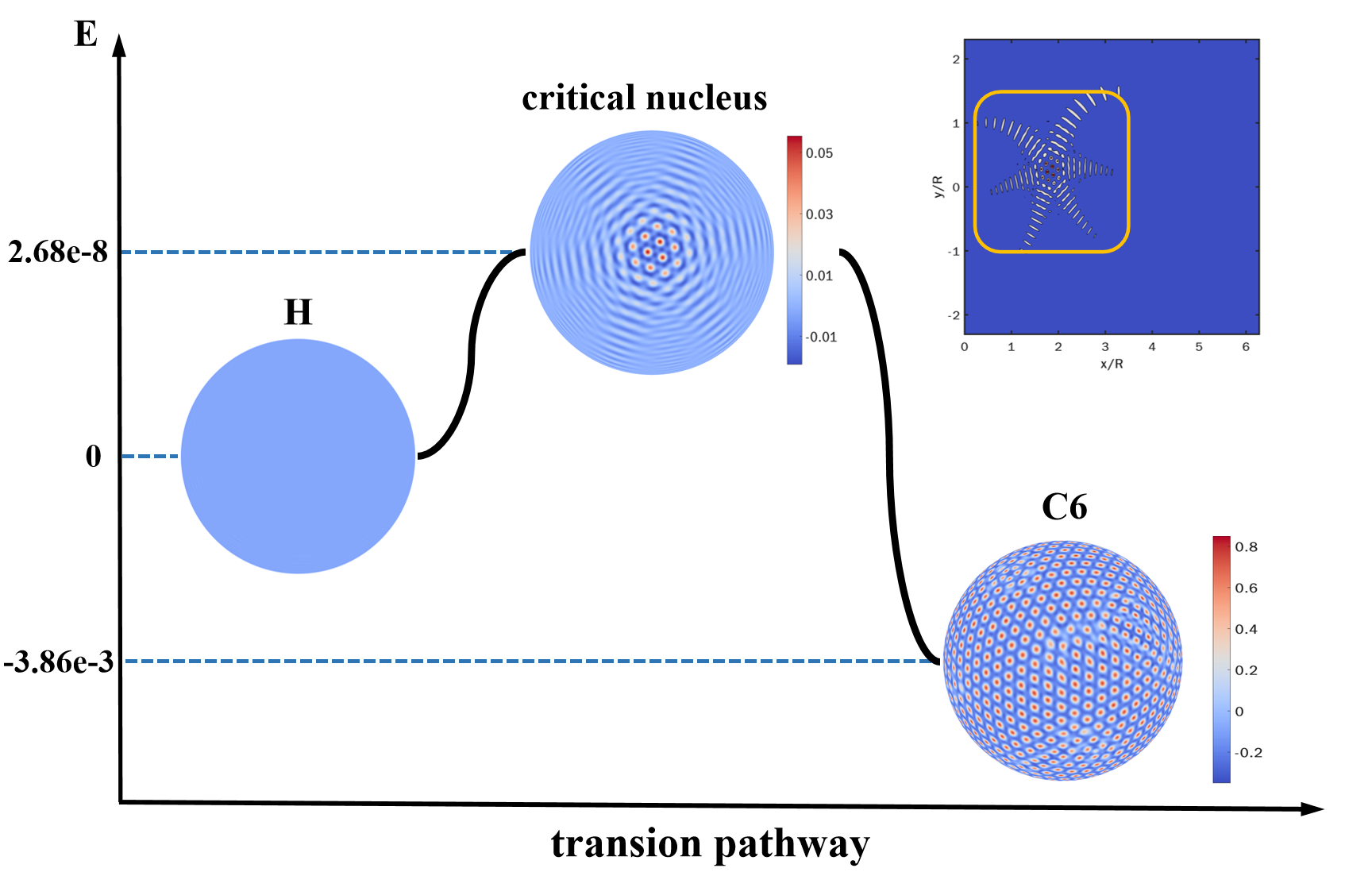}
	\caption{Transition path from H to spherical C6, computed using the NPSS method for the spherical LP model with $\alpha=-0.002$, $\gamma=0.8$, and $R=57$. The inset shows the Miller cylindrical projection of the index-1 saddle point.}
	\label{fig:H_C6}
\end{figure}

Using the NPSS method, we compute the minimum energy path connecting H and C6
and identify a transition state with an activation barrier
$\Delta E=2.68\times 10^{-8}$, as shown in Fig.~\ref{fig:H_C6}. For
visualization, the inset in Fig. 3 shows a Miller cylindrical projection of the critical nucleus.
The transition state corresponds to a localized C6 nucleus
embedded in a weakly modulated background. Around the nucleus, faint
stripe-like density waves extend along the principal lattice directions,
indicating the onset of sixfold crystalline ordering on the curved surface.

\begin{figure*}[!htbp]
 \centering
 \includegraphics[width=0.8\textwidth]{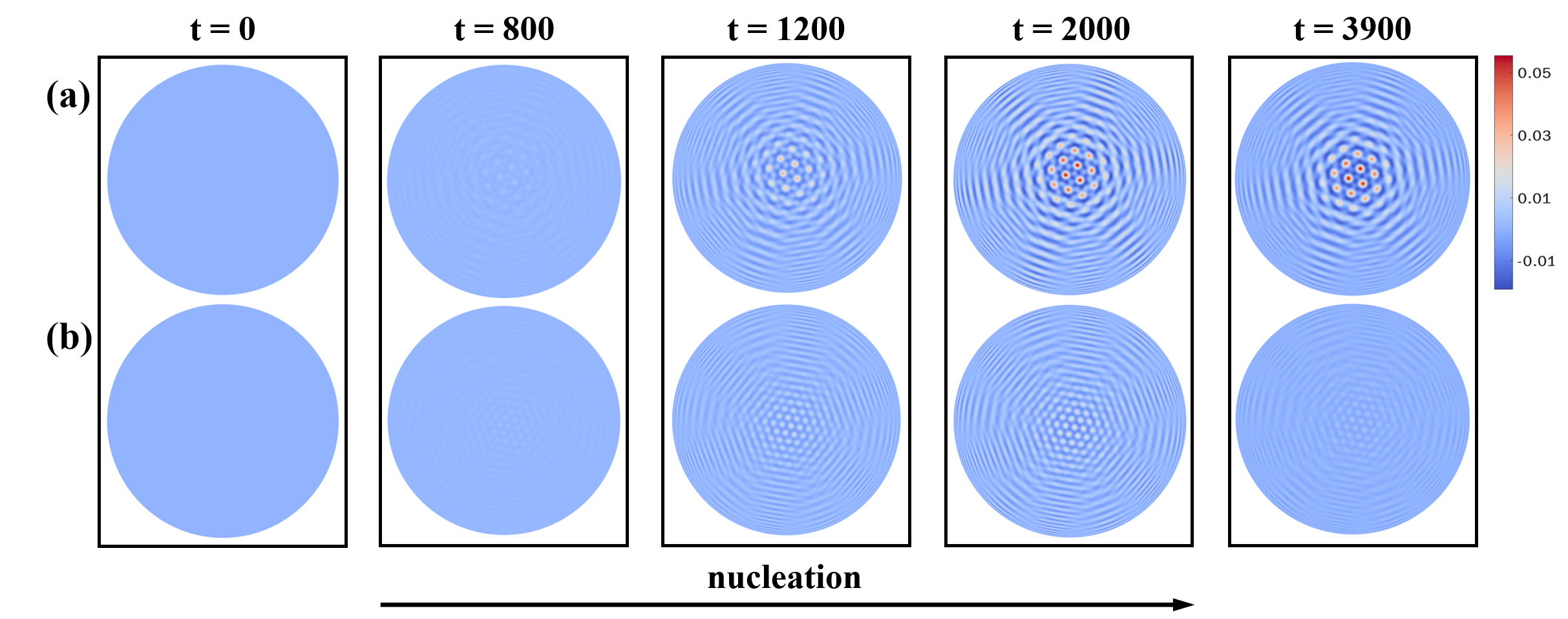}
 \caption{Early-stage C6 nucleation along the transition path in
 Fig.~\ref{fig:H_C6}. (a) Front-side nucleation region, where a localized C6
 embryo develops from weak density fluctuations. (b) Antipodal region, where
 transient fluctuations are suppressed as the critical nucleus forms.}
 \label{fig:C6_nucleation}
\end{figure*}

The early stage of the MEP reveals how the critical nucleus emerges from weak density modulations on the closed spherical surface. As shown in
Fig.~\ref{fig:C6_nucleation}, weak density modulations initially appear over the
sphere, including both the front-side nucleation region and its antipodal
region. As the system approaches the transition state, the front-side embryo
is amplified into a well-defined C6 critical nucleus, while the antipodal
fluctuations are suppressed and return to a nearly homogeneous background.
This behavior indicates that the H-to-C6 transition is governed by localized
nucleation rather than a spatially uniform instability.

\begin{figure*}[htpb]
\centering
\includegraphics[width=0.8\textwidth]{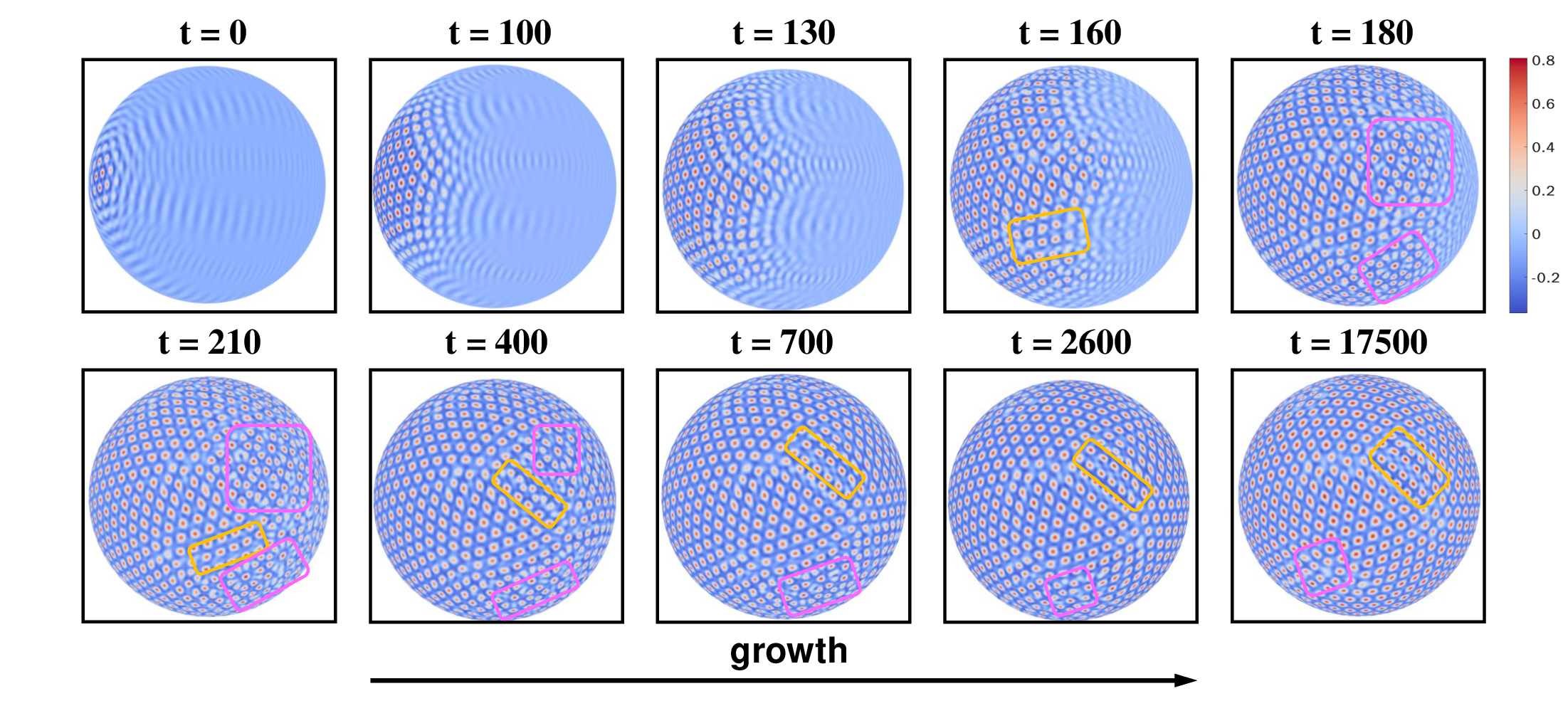}
\caption{Growth of the spherical C6 phase. The initially compact nucleus develops into branched crystalline domains, and persistent line defects remain after domain collisions.}
\label{fig:C6_growth}
\end{figure*}

After crossing the activation barrier, the C6 nucleus grows spontaneously
under the mass-conserving gradient flow, as shown in Fig.~\ref{fig:C6_growth}.
At early times, the crystalline domain expands with an approximately circular
front while preserving the local sixfold bond geometry. The frustration arises
when this locally hexagonal order is extended over the positively curved
sphere. Unlike in a flat plane, where a hexagonal lattice can maintain global
registry through translations, local lattice directions on the sphere acquire
path-dependent orientational shifts during propagation. Consequently,
different growing domains accumulate orientational and phase mismatches, which
destabilize the circular front and promote ribbon-like branched growth.

During the branched growth stage, frustrated inter-branch regions temporarily
display decagonal-like local order (highlighted by the magenta boxes), reflecting the competition between local
C6 ordering and global spherical incompatibility. As the hexagonal domains
continue to grow and collide, minor misfits can be locally relaxed, whereas
the remaining mismatch is concentrated along domain boundaries (highlighted by the orange boxes). These
boundaries eventually develop into persistent macroscopic line defects. Thus,
defect formation in spherical C6 growth results from the failure of global
lattice registry on a closed curved surface.

\subsection{Nucleation and Growth of DQC from H}\label{H_DQC}
We next examine the nucleation and growth of the spherical DQC phase from the
metastable homogeneous state H. The computation is performed on a sphere of
radius $R=57$ with $\alpha=-0.03$ and $\gamma=0.8$. For these parameter values,
H is metastable with $E=0$, whereas the relaxed DQC state has a lower
area-averaged free energy, $E=-8.74\times 10^{-4}$, providing the
thermodynamic driving force for the H-to-DQC transition.

\begin{figure}[htbp]
	\centering
	\includegraphics[width=0.48\textwidth]{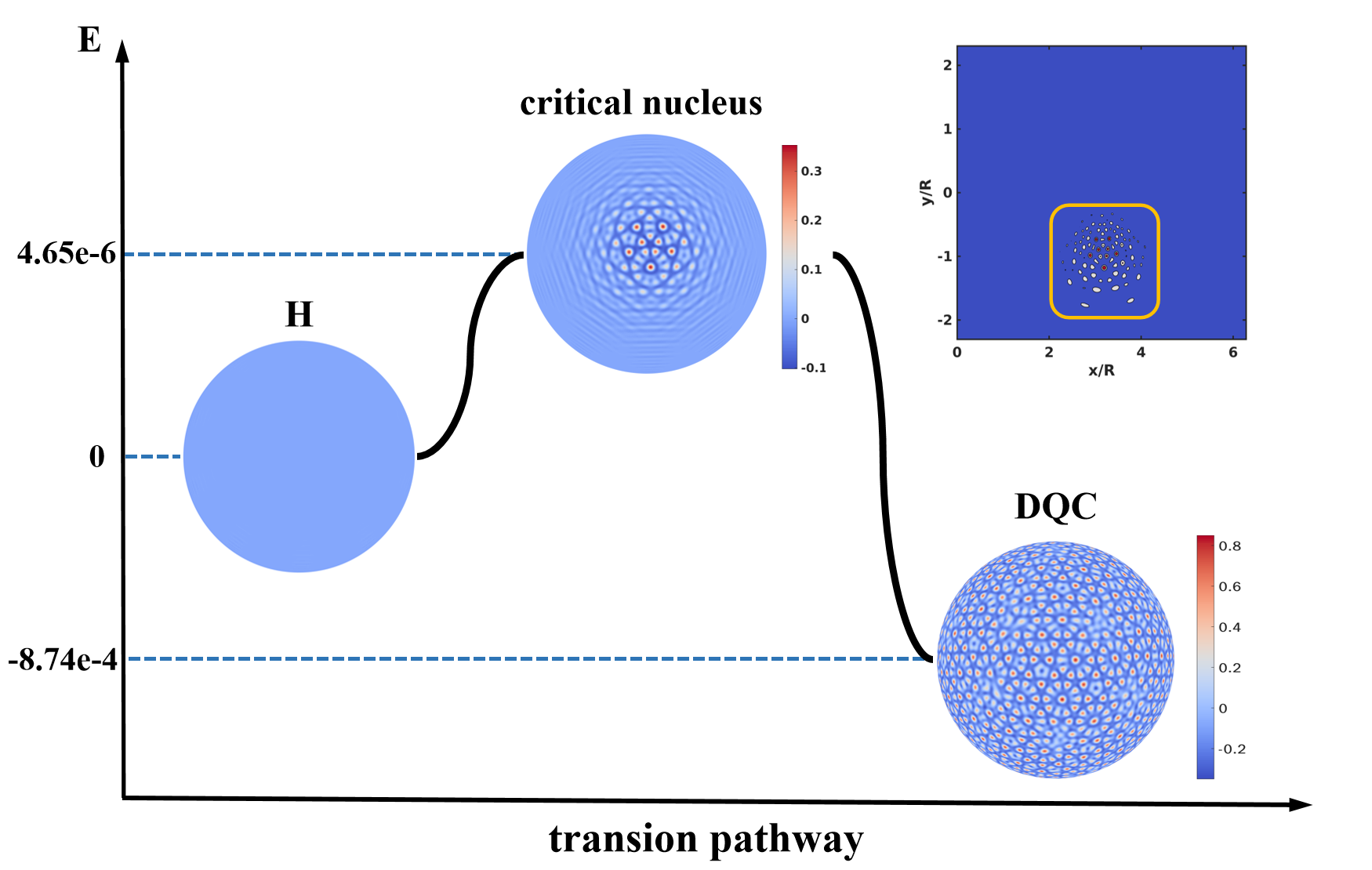}
	\caption{Transition path from H to spherical DQC computed using the NPSS
	method for the spherical LP model with $\alpha=-0.03$, $\gamma=0.8$,
	and $R=57$. The inset shows the Miller cylindrical projection of the
	index-1 saddle point.}
	\label{fig:H_DQC}
\end{figure}

Using the NPSS method, we compute the minimum energy path connecting H and
DQC and identify a transition state with an activation barrier
$\Delta E=4.65\times 10^{-6}$, as shown in Fig.~\ref{fig:H_DQC}. The
transition state is a localized DQC critical nucleus with an overall fivefold
arrangement, consisting of five coupled decagonal motifs.

\begin{figure*}[!htpb]
	\centering
	\includegraphics[width=0.8\textwidth]{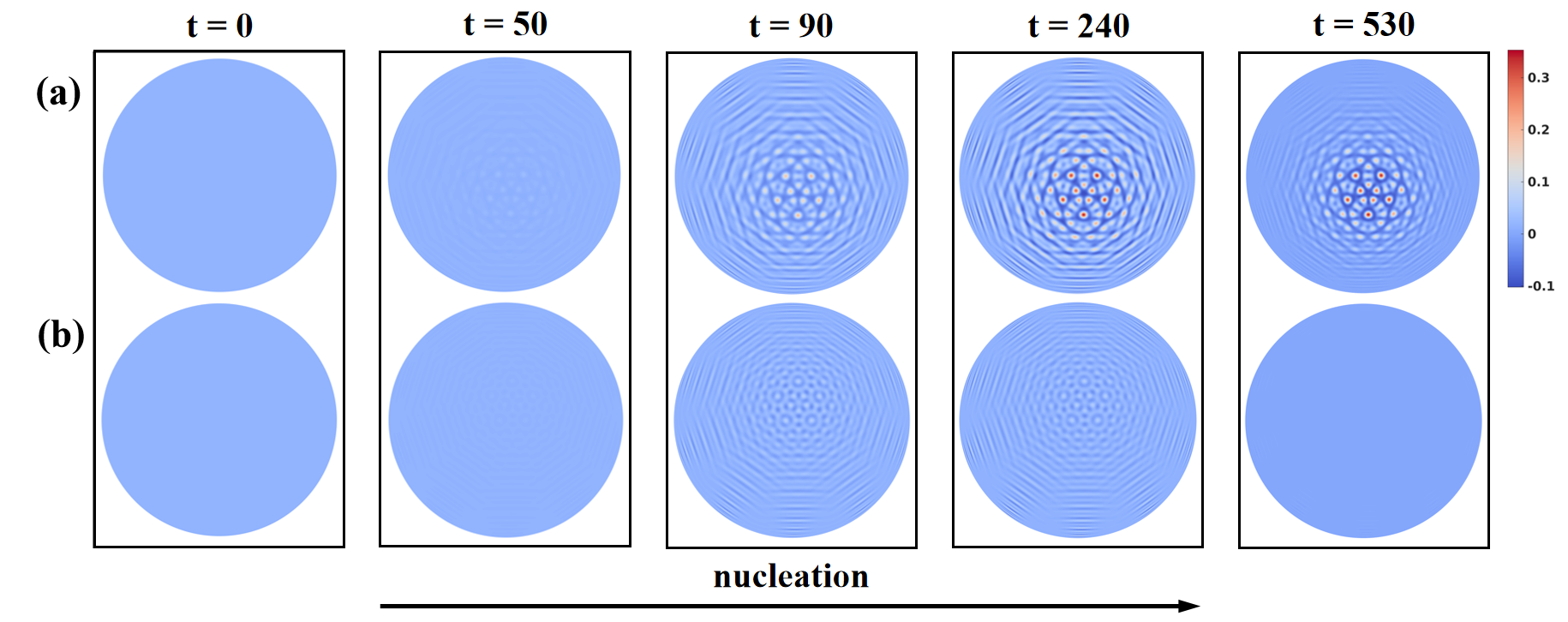}
	\caption{Nucleation of spherical DQC along the transition path. (a) Front-side
	region with a localized DQC embryo. (b) Antipodal region with weak transient
	fluctuations.}
	\label{fig:DQC_nucleation}
\end{figure*}

The early stage of the MEP shows that the DQC critical nucleus emerges from
weak density fluctuations on the spherical surface (Fig.~\ref{fig:DQC_nucleation}(a)),
while antipodal fluctuations are suppressed (Fig.~\ref{fig:DQC_nucleation}(b)). As the system
approaches the transition state, the front-side embryo develops into a well-defined DQC
nucleus. This indicates that, like the H-to-C6 transition, the H-to-DQC
transition is governed by localized nucleation rather than a uniform instability.

\begin{figure*}[!htpb]
	\centering
	\includegraphics[width=0.8\textwidth]{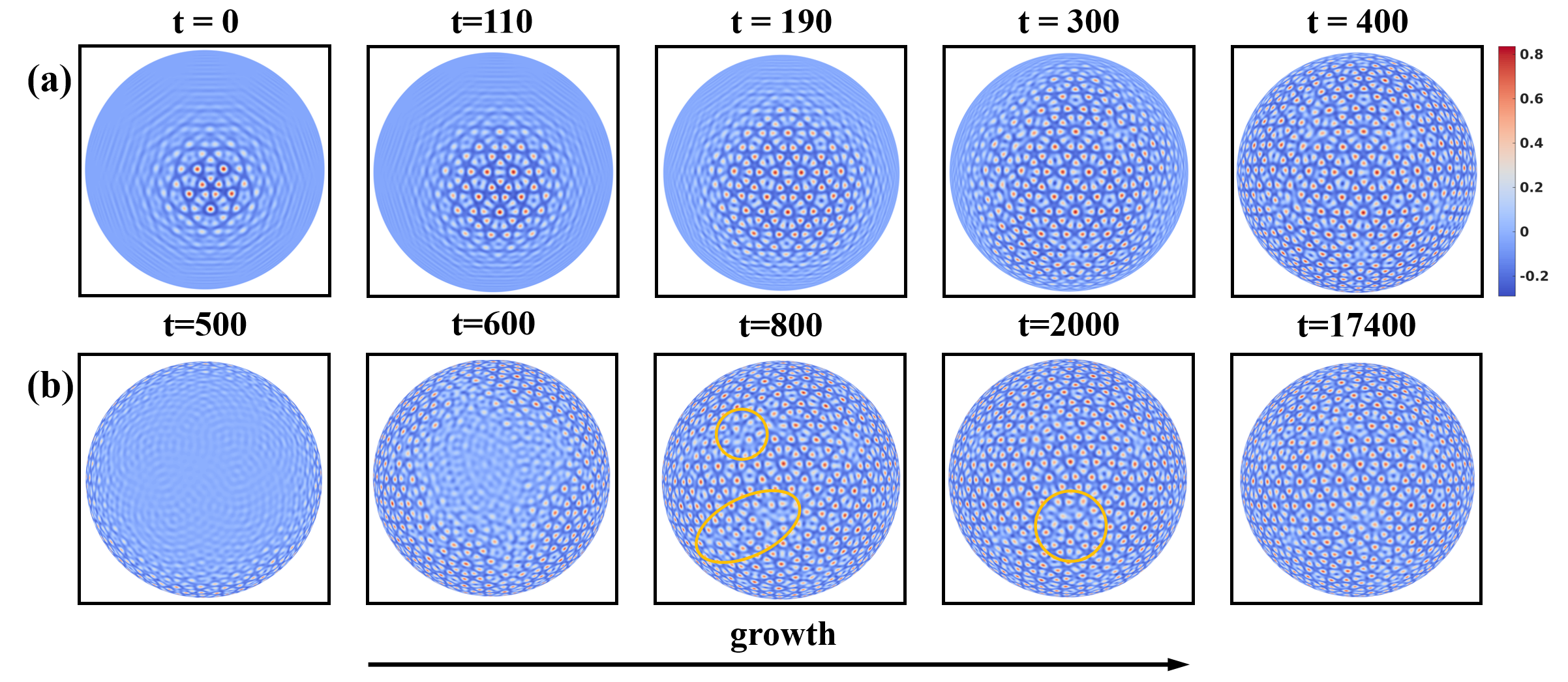}
	\caption{Growth of the spherical DQC along the transition path. Transient
	defects appear when growth fronts collide, and are gradually removed through
	local coordination rearrangements.}
	\label{fig:DQC_growth}
\end{figure*}

After crossing the nucleation barrier, the DQC nucleus grows under the
mass-conserving gradient flow (Fig.~\ref{fig:DQC_growth}). The DQC growth front remains nearly circular at early times and does not develop the pronounced branching observed for C6. The
quasiperiodic structure contains multiple local coordination motifs, allowing
the front to accommodate curvature-induced mismatch without forming persistent
branched defects. Transient line defects may appear when fronts collide near
the antipodal region, but they are gradually eliminated through local
structural rearrangements, similar to phason-like relaxation in planar
quasicrystals.

To quantify the difference between the growth of C6 and DQC, we compute the spatially averaged local bond-orientational order for extracted peak positions,
\begin{equation}
\left\langle |\psi_n| \right\rangle
=
\frac{1}{N}\sum_{i=1}^{N}|\psi_n(i)|,
\end{equation}
where $n=6$ for C6 and $n=10$ for DQC, and $N$ is the number of peaks
(see Appendix~\ref{app:M_Order}). Fig.~\ref{fig:Global_order} separates growth into domain expansion
($A_i$--$B_i$) and subsequent structural relaxation
($B_i$--$C_i$), with $i=1$ for C6 and $i=2$ for DQC.
For C6, the mean local sixfold order decreases markedly during
domain expansion and recovers as the domains reorganize,
consistent with the branched growth and persistent boundary
defects shown in Fig.~\ref{fig:C6_growth}.
For DQC, the corresponding tenfold order varies more moderately
and approaches a plateau as transient defects relax, consistent
with the smoother growth morphology in
Fig.~\ref{fig:DQC_growth}.
Together, the structural snapshots and bond-order evolution
distinguish the two modes of growth and relaxation.
The comparison concerns the evolution along each trajectory,
since the two simulations use different parameter values
and symmetry indices.

\begin{figure}[htbp]
 \centering
 \includegraphics[width=0.48\textwidth]{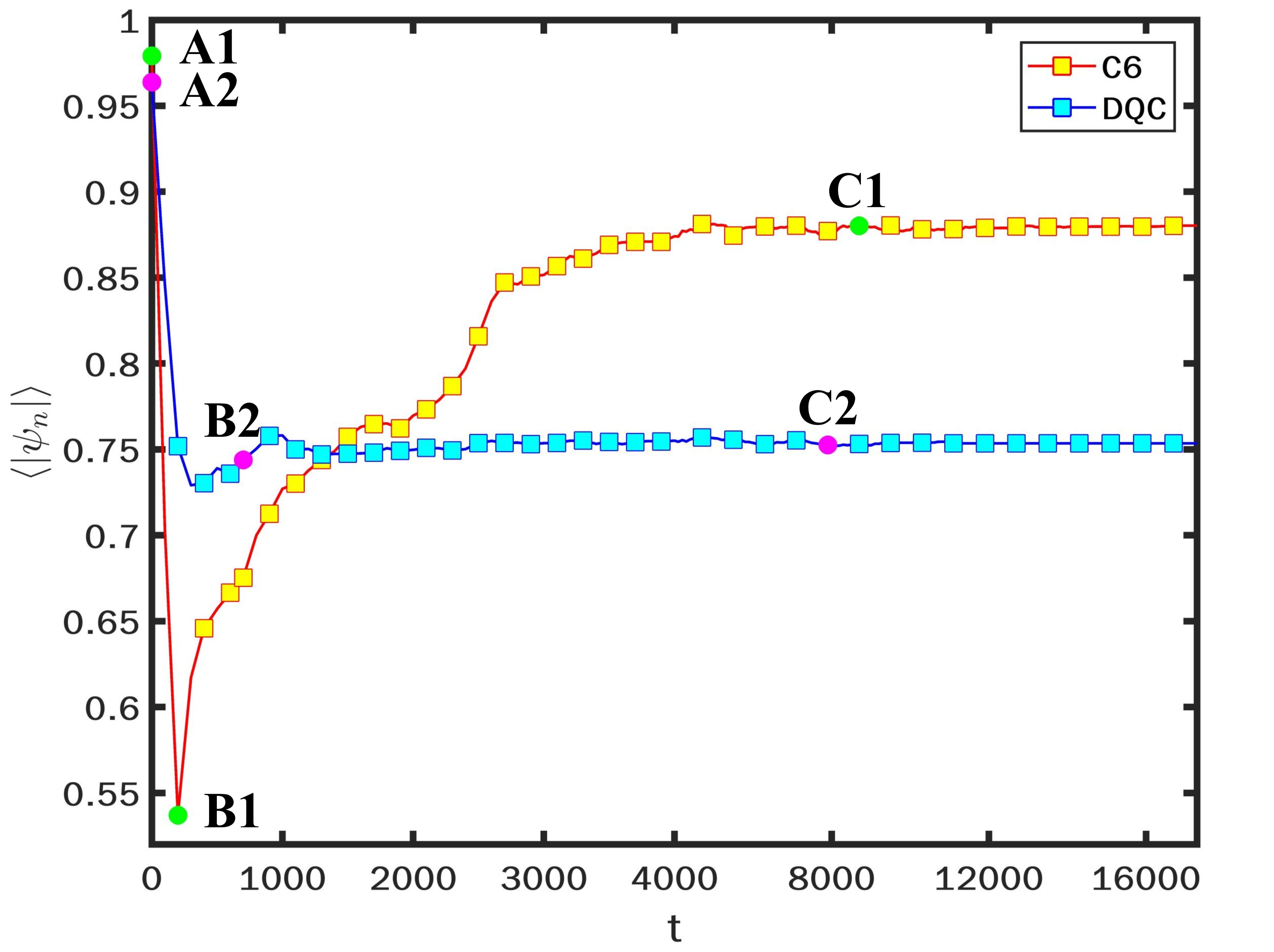}
 \caption{Spatially averaged local bond-orientational order $\langle |\psi_n|\rangle$ during the growth of C6 ($n=6$) and DQC ($n=10$). The two curves correspond to the respective parameter values used in Figs.~\ref{fig:C6_growth} and \ref{fig:DQC_growth}. Therefore, their temporal trends, rather than their absolute magnitudes, are compared. The markers $A_i$, $B_i$, and $C_i$ indicate the onset of growth, full coverage of the spherical surface by ordered domains, and the completion of defect relaxation, respectively, with $i=1$ for C6 and $i=2$ for DQC.}
 \label{fig:Global_order}
\end{figure}

\subsection{Phase Transition from DQC to C6}
\label{DQCtoC6}
\begin{figure*}[!htpb]
    \centering
    \includegraphics[width=0.95\textwidth]
    {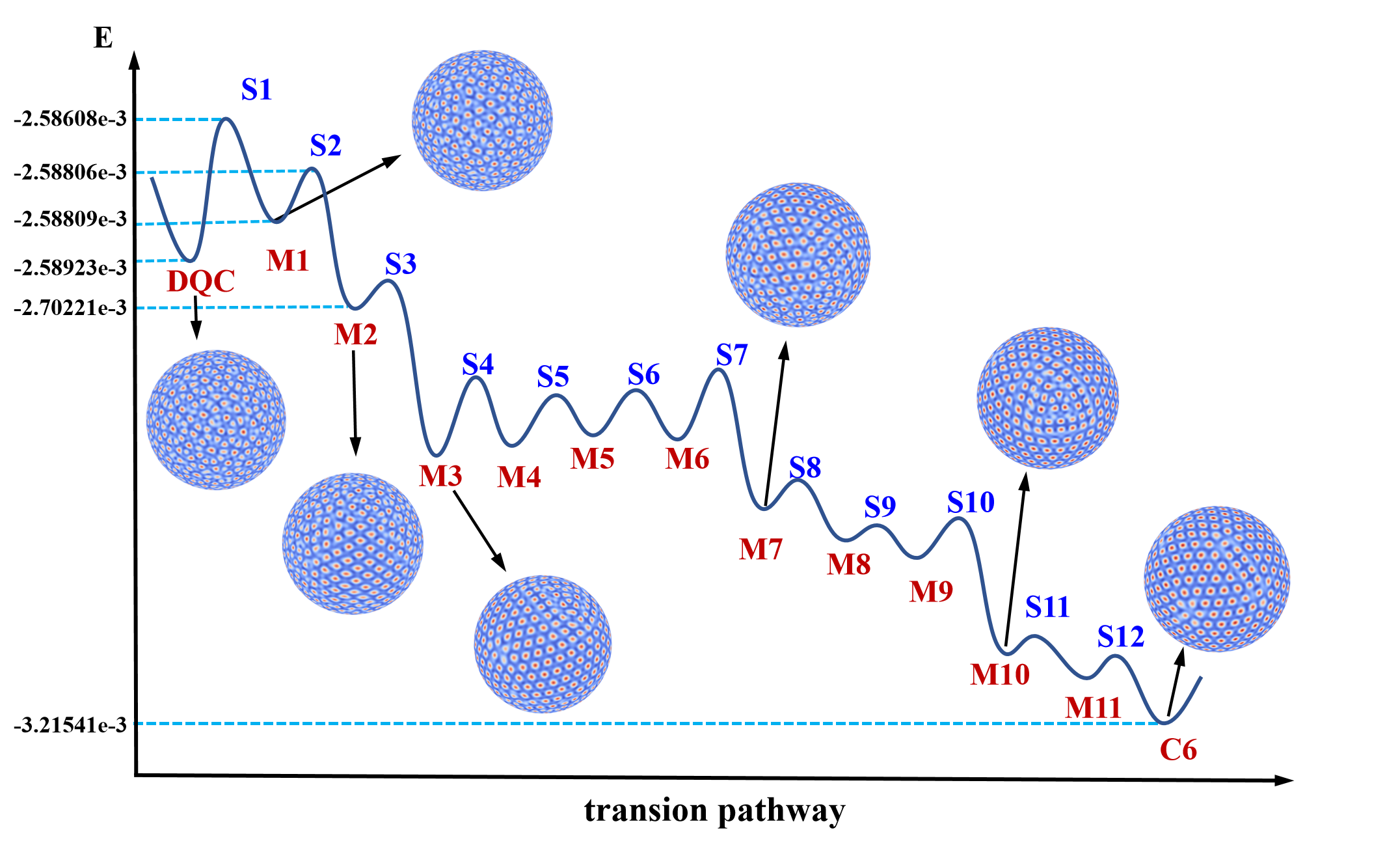}
    \caption{Multi-step transition pathway from DQC to C6
    computed using the NPSS method for the spherical LP model
    with $\alpha=-0.006$, $\gamma=0.8$, and $R=38.5$.
    The pathway contains twelve transition states and eleven
    metastable intermediates. The insets show representative
    configurations during local structural conversion and
    domain-boundary reorganization.}
    \label{fig:DQC_C6}
\end{figure*}

We compute the transition pathway from spherical DQC to C6
at $R=38.5$, $\alpha=-0.006$, and $\gamma=0.8$.
As shown in Fig.~\ref{fig:DQC_C6}, the pathway contains twelve
transition states and eleven metastable intermediates.
The first transition produces $M_1$, whose free energy is
slightly higher than that of the initial DQC.
Further local reconstruction of decagonal motifs leads to
$M_2$, followed by pronounced C6 domain growth and a substantial
energy decrease between $M_2$ and $M_3$.
During the later stages, C6 domains progressively replace
the remaining DQC regions, accompanied by domain-boundary
reorganization. Persistent line defects remain in the final
C6 structure.

\begin{figure}[!htpb]
    \centering
    \includegraphics[width=0.48\textwidth]
    {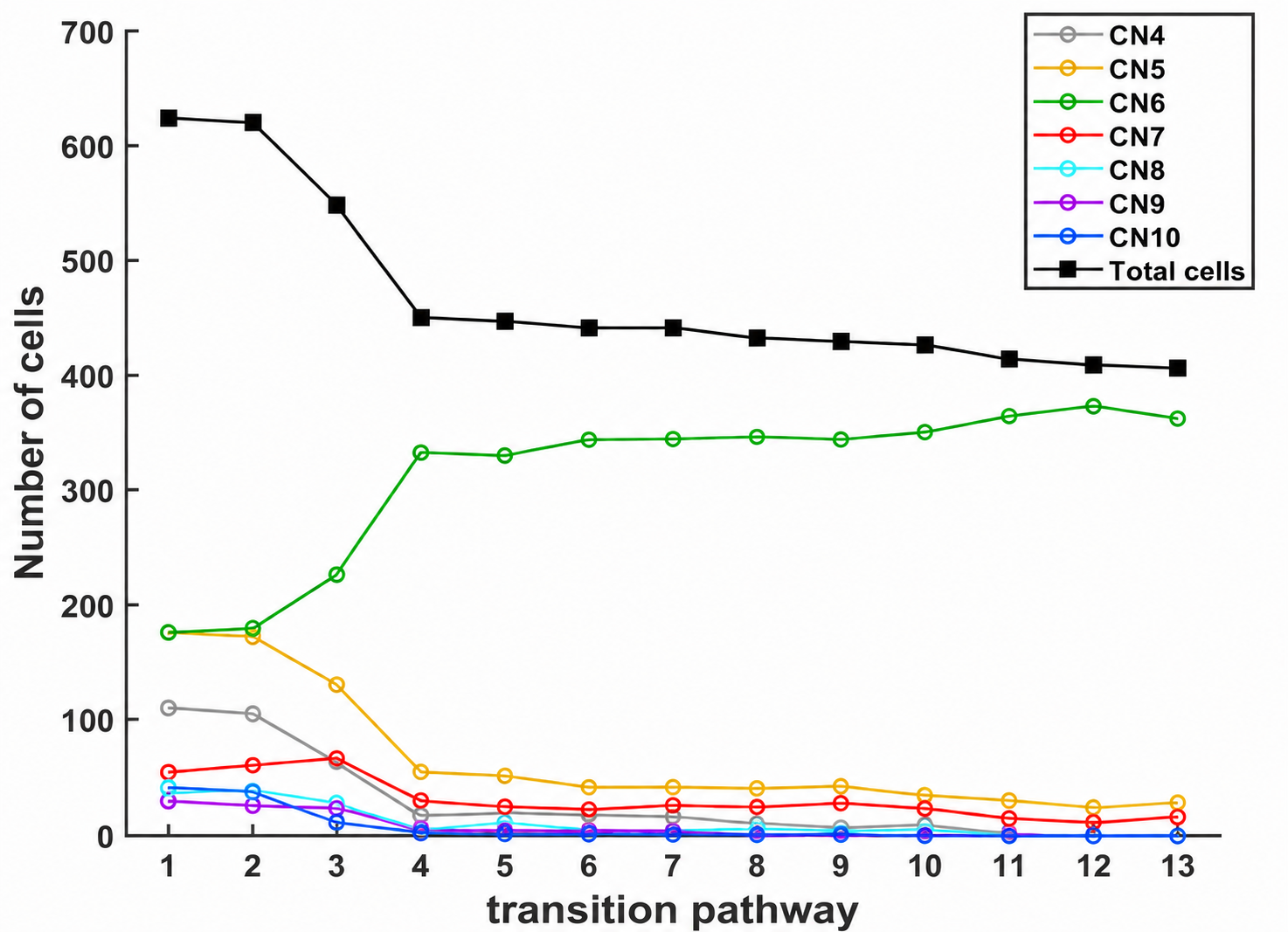}
    \caption{Evolution of Voronoi coordination populations
    along the DQC-to-C6 pathway in Fig.~\ref{fig:DQC_C6}.
    Positions 1 and 13 correspond to DQC and C6, respectively,
    and positions 2--12 correspond to $M_1$--$M_{11}$.
    The number of CN6 cells increases as C6 domains develop,
    while the total number of density peaks decreases.
    CN5 and CN7 cells persist in the final structure.}
    \label{fig:voron_cell_number}
\end{figure}

To characterize these structural changes, we analyze the
spherical Voronoi network constructed from the density peaks.
Figure~\ref{fig:voron_cell_number} shows the populations of
different coordination environments, while
Fig.~\ref{fig:vorons} displays their spatial distributions.
Here, CN$i$ denotes a Voronoi cell with $i$ neighbors.
As the transformation proceeds, the number of CN6 cells
increases, whereas the total number of density peaks decreases.
The diverse coordination environments of the parent DQC
are progressively replaced by predominantly sixfold coordination.
Small populations of highly coordinated cells persist in
residual DQC regions during intermediate stages, reflecting
the spatially heterogeneous conversion of the parent network.
In the final structure, the remaining non-sixfold cells
are predominantly CN5 and CN7 and are concentrated near
C6 domain boundaries.

These observations show that increasing local crystalline
order is accompanied by the localization of residual
coordination mismatch.
Non-sixfold environments form part of the characteristic
local order of the parent DQC, whereas the product C6
develops predominantly sixfold-coordinated domain interiors.
The remaining CN5 and CN7 cells accommodate mismatch near
their boundaries.
Within a locally triangular crystal, these cells correspond
to positive and negative disclinations, and appropriately
arranged five--seven pairs can form dislocations and
grain-boundary structures~\cite{bausch2003grain,bowick2006crystalline}.
Meanwhile, the decrease in the total number of density peaks
indicates that the transformation reorganizes both the
peak population and its connectivity.
This is compatible with the zero-mean constraint, which
conserves the integral of the density variation rather than
the number of density maxima.

Spherical topology constrains this reconstruction.
For a triangulated spherical Delaunay network, Euler's
theorem gives
\begin{equation}
\sum_i (6-i)N_i=12,
\label{eq:euler_coordination_constraint}
\end{equation}
where $N_i$ is the number of cells with $i$ neighbors.
Once only CN5, CN6, and CN7 cells remain, this relation
reduces to $N_5-N_7=12$.
Thus, the local coordination environments and their spatial
distribution evolve during the transformation, while the
net coordination charge remains fixed by spherical topology.

The observed reconstruction connects the DQC-to-C6 pathway
to the distinct modes of mismatch accommodation identified
during growth from H.
DQC relaxes collision-induced defects through local
coordination rearrangements, whereas C6 retains residual
orientational and phase mismatches as persistent line defects.
During the solid-to-solid transformation, the parent DQC
network is progressively reconstructed into locally hexagonal
domains, while residual mismatch becomes concentrated near
their boundaries.
The structural transition therefore converts local
rearrangement within the quasicrystalline network into
boundary-defect accommodation in the product crystal.

This conversion involves coupled adjustments of local order
and interdomain compatibility.
Establishing sixfold coordination within an emerging domain
does not by itself resolve its mismatch with neighboring
ordered regions.
Local network reconstruction must therefore be accompanied
by domain-boundary reorganization.
The successive metastable intermediates resolve these
adjustments, linking the multi-step character of the computed
pathway to the different structural requirements for
accommodating mismatch in DQC and C6.
Together with the two homogeneous-to-ordered pathways,
this result shows how local structure governs both the
accommodation of spherical frustration and the conversion
between distinct modes of defect organization.
Additional views of the intermediates and their Voronoi
diagrams are provided in Appendix~\ref{app:stable_states}.

\section{Conclusions}

We investigate the nucleation, growth, and solid-to-solid
transformation of spherical C6 and DQC using the
Lifshitz--Petrich model. Combining a spherical harmonic
pseudospectral method with the nullspace-preserving saddle
search method, we obtain the H-to-C6, H-to-DQC, and DQC-to-C6 pathways, and reveal distinct modes of geometric-mismatch
accommodation. Both homogeneous-to-ordered transitions pass
through localized critical nuclei, but C6 growth produces
branched domains and persistent line defects, whereas DQC
relaxes collision-induced defects through local coordination
rearrangements. The DQC-to-C6 transition connects these two modes through
a sequence of metastable intermediates. Their different
structural requirements couple the development of local
sixfold order to the adjustment of compatibility between
emerging domains. As C6 domains replace the parent DQC
network, residual non-sixfold coordination becomes concentrated
near domain boundaries, while the net coordination charge
remains constrained by spherical topology. The multi-step character of the computed pathway reflects
the progressive conversion from local rearrangement within
the quasicrystalline network to boundary-defect accommodation
in the product crystal. These findings connect local
structural order, mismatch accommodation, and multi-step
phase transformation on a spherical surface.

\begin{acknowledgments}
This work was supported by the National Key R\&D Program of China (2023YFA1008802), the Science and Technology Innovation Program of Hunan Province (2024RC1052), and the Innovative Research Group Project of National Natural Science Foundation of Hunan Province of China (2024JJ1008). We acknowledge the High-Performance Computing Platform of Xiangtan University for providing computational resources. We are grateful to Dr. Qun Qiu for her assistance with the algorithms and code.
\end{acknowledgments}

\appendix
\section{Spherical Harmonic Pseudospectral Discretization}
\label{app:discretization}

The spherical LP free energy in Eq.~\eqref{eq:2-1} is discretized by a
truncated spherical harmonic expansion,
\begin{equation}
\psi_N(\theta,\varphi)
=
\sum_{\ell=0}^{N}\sum_{m=-\ell}^{\ell}
\hat{\psi}_{\ell}^{m}Y_{\ell}^{m}(\theta,\varphi),
\label{eq:expansion}
\end{equation}
where $\hat{\psi}_{\ell}^{m}$ are the spectral coefficients and
$Y_{\ell}^{m}$ are spherical harmonics normalized with respect to the
area-averaged inner product:
\begin{equation}
\frac{1}{|\mathbb{S}^2_R|}
\int_{\mathbb{S}^2_R}
Y_{\ell}^{m}\overline{Y_{\ell'}^{m'}}\,\mathrm{d}s
=
\delta_{\ell\ell'}\delta_{mm'}.
\label{eq:sh_orthogonality}
\end{equation}
With the Condon--Shortley phase convention, the coefficients of the
real-valued field $\psi_N$ satisfy
$\hat{\psi}_{\ell}^{m} =
(-1)^m\overline{\hat{\psi}_{\ell}^{-m}}$.
In the computations, we use the truncation degree $N=255$.

The Laplace--Beltrami operator is diagonal in the spherical harmonic basis:
\begin{equation}
\Delta_{\mathbb{S}^2_R}Y_{\ell}^{m}
=
-\frac{\ell(\ell+1)}{R^2}Y_{\ell}^{m}.
\end{equation}
Thus, the linear differential operator is evaluated exactly within the
truncated spectral space. By orthogonality, the quadratic interaction
contribution to the area-averaged free energy becomes
\begin{equation}
G_N(\hat{\Psi})
=
\frac{1}{2}
\sum_{\ell=0}^{N}\sum_{m=-\ell}^{\ell}
\left(1-\frac{\ell(\ell+1)}{R^2}\right)^2
\left(q^2-\frac{\ell(\ell+1)}{R^2}\right)^2
\left|\hat{\psi}_{\ell}^{m}\right|^2,
\label{eq:GN}
\end{equation}
where $\hat{\Psi}$ denotes the vector of all retained spectral coefficients.

The local potential contribution is evaluated by a pseudospectral
procedure. The field $\psi_N$ is transformed to a grid with
$N_\theta=256$ Gauss--Legendre nodes in $\cos\theta$ and
$N_\varphi=512$ uniformly spaced longitude points. The polynomial terms
are computed pointwise and integrated by quadrature:
\begin{equation}
F_N(\hat{\Psi})
=
\frac{1}{|\mathbb{S}^2_R|}
\sum_{a=1}^{N_\theta}\sum_{b=1}^{N_\varphi}
\left(
-\frac{\alpha}{2}\psi_{ab}^2
-\frac{\gamma}{3}\psi_{ab}^3
+\frac{1}{4}\psi_{ab}^4
\right)
w_a\,\Delta\varphi,
\label{eq:FN}
\end{equation}
where $\psi_{ab}=\psi_N(\theta_a,\varphi_b)$,
$\Delta\varphi=2\pi/N_\varphi$, and $w_a=R^2\omega_a$, with
$\omega_a$ denoting the standard Gauss--Legendre weights on $[-1,1]$.
The forward and inverse spherical harmonic transforms are performed
using SHTns~\cite{schaeffer2013efficient}. This pseudospectral evaluation
avoids the explicit assembly of high-order spectral convolution tensors.

The discrete constrained minimization problem is therefore
\begin{equation}
\min_{\hat{\Psi}} E_N(\hat{\Psi})
=
G_N(\hat{\Psi})+F_N(\hat{\Psi}),
\quad
\text{s.t.}\quad \hat{\psi}_{0}^{0}=0.
\label{eq:discrete_optimization}
\end{equation}
The constraint $\hat{\psi}_{0}^{0}=0$ enforces the zero-mean condition
throughout the relaxation and saddle-search computations.

\section{Bond-Orientational Order}
\label{app:M_Order}

To quantify local structural order, discrete peak positions
$\mathcal{V}_{N_{\mathrm p}}
=\{\mathbf{r}_1,\ldots,\mathbf{r}_{N_{\mathrm p}}\}
\subset\mathbb{S}^2_R$
are extracted from the local maxima of the continuous density field
$\psi(\mathbf{r})$. The neighbor set $\mathcal{N}(\mathbf{r}_i)$ of each peak
is determined by spherical Voronoi tessellation~\cite{boots1999spatial}.

For each peak $\mathbf{r}_i$, we construct a local tangent plane perpendicular
to the radial normal $\mathbf{n}_i=\mathbf{r}_i/R$. The bond vector from
$\mathbf{r}_i$ to a neighboring peak $\mathbf{r}_j$ is projected onto this
tangent plane, and its polar angle is denoted by $\phi_{ij}$. The local
$n$-fold bond-orientational order parameter is then defined as
\begin{equation}
\psi_n(i)
=
\frac{1}{|\mathcal{N}(\mathbf{r}_i)|}
\sum_{j\in\mathcal{N}(\mathbf{r}_i)}
e^{\mathrm{i}n\phi_{ij}},
\label{eq:local_psi_n}
\end{equation}
following the standard bond-order construction~\cite{steinhardt1983bond}.
Although $\phi_{ij}$ depends on the choice of the local tangent-plane basis,
the magnitude $|\psi_n(i)|$ is invariant under a rotation of that basis.

The spatially averaged local bond-orientational order used in the main text
is defined as
\begin{equation}
\left\langle |\psi_n| \right\rangle
=
\frac{1}{N_{\mathrm p}}
\sum_{i=1}^{N_{\mathrm p}}|\psi_n(i)|.
\end{equation}
This quantity measures the mean strength of local $n$-fold
bond-orientational order over the sphere. Because the complex phase of
$\psi_n(i)$ is removed before averaging, it does not measure the global
orientational coherence between spatially separated domains.

\section{Metastable Intermediates and Voronoi Diagrams}
\label{app:stable_states}

\begin{figure*}[!htpb]
	\centering
	\includegraphics[width=0.95\textwidth]{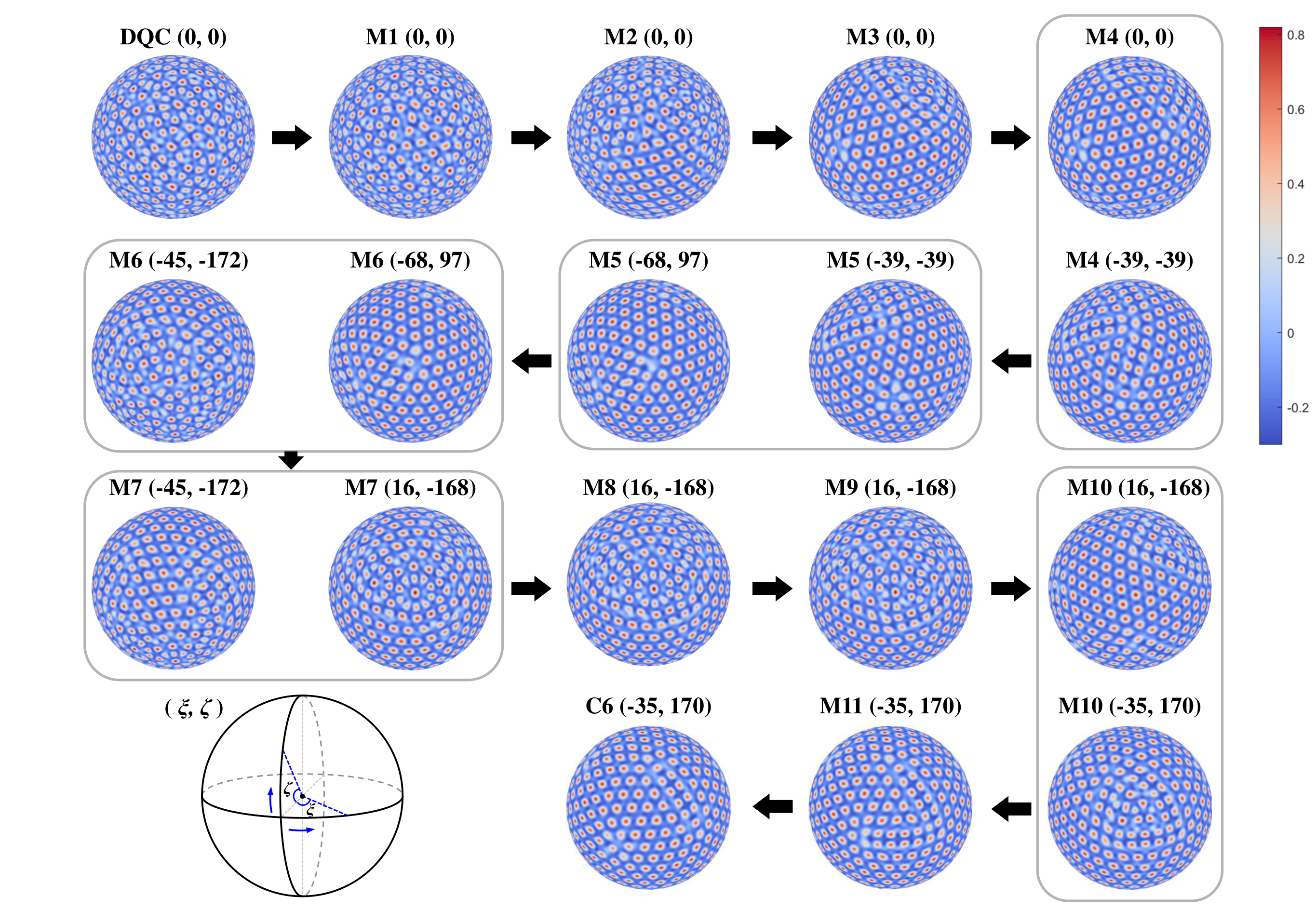}
	\caption{Metastable intermediates along the DQC-to-C6 transition pathway.
	The states $M_i$ ($i=1,2,\ldots,11$) correspond to the local minima
	resolved on the MEP in Fig.~\ref{fig:DQC_C6}. Multiple viewing angles are
	shown to visualize the heterogeneous structural reorganization on the
	sphere. Rounded boxes group different views of the same configuration.}
	\label{fig:minimums}
\end{figure*}

This appendix provides additional views of the metastable intermediates along
the DQC-to-C6 pathway discussed in Sec.~\ref{DQCtoC6}. The configurations
$M_i$ ($i=1,2,\ldots,11$) correspond to the local minima shown in
Fig.~\ref{fig:DQC_C6}. Figure~\ref{fig:minimums} displays selected
intermediate states from multiple viewing angles, showing the heterogeneous
breakdown of the DQC and the subsequent growth of C6 domains.

\begin{figure*}[!htpb]
	\centering
	\includegraphics[width=0.95\textwidth]{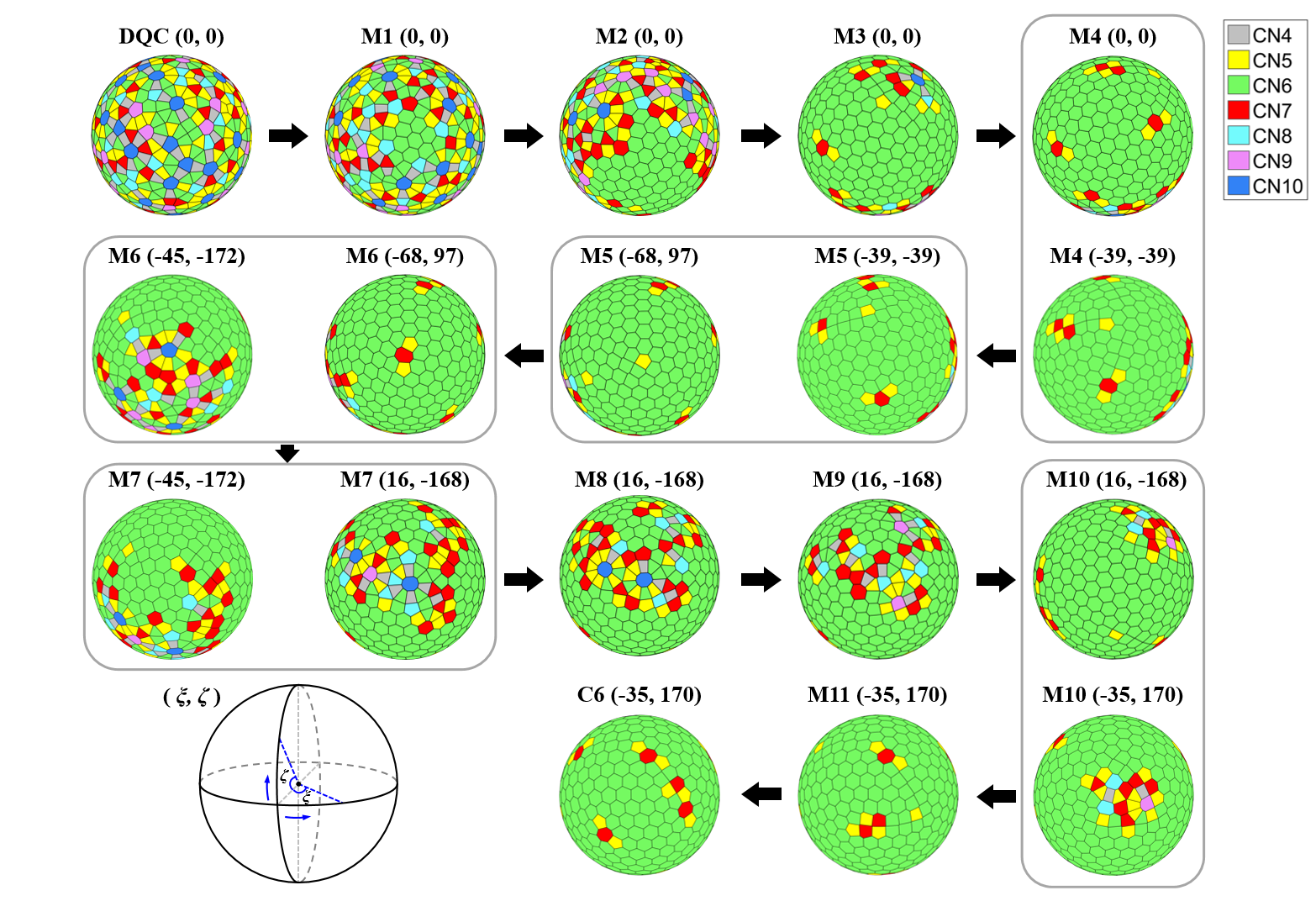}
	\caption{Spherical Voronoi diagrams corresponding to the metastable
	intermediates shown in Fig.~\ref{fig:minimums}. Here, $\mathrm{CN}_i$
	($i=4,5,\ldots,10$) denotes a Voronoi cell with coordination number $i$.}
	\label{fig:vorons}
\end{figure*}

The viewing orientation of each configuration is specified by the pair
$(\xi,\zeta)$ shown above the corresponding panel, where
$\xi\in[-180^{\circ},180^{\circ})$ denotes the rotation in the meridional
direction and $\zeta\in[-180^{\circ},180^{\circ})$ denotes the rotation in the
latitudinal direction. The lower-left schematic in Fig.~\ref{fig:minimums}
defines the positive directions of these two rotations. The same viewing
orientations are used for the density fields in Fig.~\ref{fig:minimums} and
the corresponding Voronoi diagrams in Fig.~\ref{fig:vorons}.

Figure~\ref{fig:vorons} shows the corresponding spherical Voronoi diagrams.
The early intermediates contain a broad distribution of coordination numbers,
whereas later states are dominated by six-coordinated cells. The remaining
non-six-coordinated cells are mainly concentrated near grain boundaries in
the final C6 state.

\bibliography{reference}

\end{document}